\documentclass[doublecol]{epl2}

\usepackage{hyperref}
\hypersetup{
breaklinks=true,
colorlinks=true,
citecolor=blue,
linkcolor=blue,
filecolor=blue,
urlcolor=blue
}

\usepackage{ulem}

\title{Protecting clean critical points by local disorder correlations}

\author{Jos\'e A. Hoyos\inst{1} \and Nicolas Laflorencie\inst{2} \and Andr\'e P. Vieira\inst{3} \and Thomas Vojta\inst{4}}
\shortauthor{J. A. Hoyos, N. Laflorencie, A. P. Vieira and T. Vojta}

\institute{
	\inst{1} Instituto de F\'{\i}sica de S\~ao Carlos, Universidade de S\~ao Paulo, C.P. 369, S\~ao Carlos, S\~ao Paulo 13560-970, Brazil\\
	\inst{2} Laboratoire de Physique des Solides, Universit\'e Paris-Sud, UMR-8502 CNRS, 91405 Orsay, France, EU\\
	\inst{3} Instituto de F\'{\i}sica, Universidade de S\~ao Paulo, C.P. 66318, S\~ao Paulo 05314-970, Brazil\\
	\inst{4} Department of Physics, Missouri University of Science and Technology, Rolla, Missouri 65409, USA
}


\abstract{
We show that a broad class of quantum critical points can be stable against {\it locally} correlated disorder even if they are unstable against uncorrelated disorder. 
Although this result seemingly contradicts the Harris criterion, it follows naturally from the absence of a random-mass term in the associated order-parameter field theory. 
We illustrate the general concept with explicit calculations for quantum spin-chain models. 
Instead of the infinite-randomness physics induced by uncorrelated disorder, we find that weak locally correlated disorder is irrelevant. 
For larger disorder, we find a line of critical points with unusual properties such as an increase of the entanglement entropy with the disorder strength. 
We also propose experimental realizations in the context of quantum magnetism and cold-atom physics. 
}

\begin{document}

\maketitle
\noindent \textit{Introduction---} The effects of quenched disorder
in condensed matter are various and constitute an interesting and
important field of research. For instance, randomness can change the
universality class of a critical point and even originate novel phases.
A paradigmatic effect of disorder near a phase transition is known
as {}``random mass:'' As there is no translational invariance, different
regions of the same system can be at different distances from 
criticality~\footnote{In field-theory language, it is as if the mass of the field were random in
  space. For thermal phase transitions, this idea is also known as
  random-$T_c$.}.
Thus, disorder fluctuations induce large and rare regions which can
be {}``locally'' in one phase while the bulk is in the other one.
This yields the so-called Griffiths singularities \cite{griffiths-prl69,mccoy-prl69,vojta-review06}.

Using the random-mass concept, Harris \cite{harris-jpc74} formulated
a simple criterion for the relevance of disorder at continuous phase
transitions: If $d\nu<2$ (with $d$ being the spatial dimension and
$\nu$ the clean correlation-length exponent), then the clean critical
behavior is destabilized by weak disorder. This is the famous Harris
criterion. It applies to the case of uncorrelated disorder. Spatial
correlations among the random masses modify the criterion. When such
correlations decay as $x^{-a}$ with distance $x$, the Harris criterion
reads $\min\{d,a\}\nu<2$ \cite{weinrib-halperin-prb83}. This implies
that correlations are relevant only when they decay slower than $x^{-d}$.
Interestingly, when the disorder correlations are relevant, the Griffiths
singularities are also enhanced \cite{rieger-correlated}. From the
above arguments, one may expect that {\it local} (i.e., short-range)
disorder correlations do not change the relevance or irrelevance of
the disorder.

In this Letter, we show that this is not always true. Specifically,
we demonstrate that certain types of local correlations can render
the disorder perturbatively {\it irrelevant} even though $d\nu<2$,
thereby stabilizing the clean critical point against weak disorder.
Although this result appears to contradict the Harris criterion, it
arises naturally when disorder correlations make the mass term spatially
uniform. If the strength of the correlated disorder is increased beyond
a critical value, where perturbative methods cannot be used anymore,
a line of finite-disorder critical points (tuned by disorder strength)
appears.\\
\textit{Field theory---} Before turning to a specific microscopic
model, let us consider the relevance of disorder within a general
field theory framework. Consider a (clean) Euclidean quantum-field
theory given by the action $S_{0}[\phi]=\int{\rm d}\tau{\rm d}{\bf x}{\cal L}_{0}[\phi]$
($\mathbf{x}$ represents position in $d$ dimensions and $\tau$
is imaginary time). The action $S_{0}$ is perturbed by a disorder
term $S_{{\rm dis}}=\sum_{{\cal O}}\int{\rm d}\tau{\rm d}\mathbf{x}\,\lambda_{{\cal O}}(\mathbf{x}){\cal O}(\mathbf{x},\tau)$.
Here, the sum is over all possible operators ${\cal O}$ to which
disorder can couple, and $\lambda_{{\cal O}}(\mathbf{x})$ is a random
variable of zero mean. (As we are considering quenched disorder, $\lambda_{{\cal O}}$
depends only on the spatial coordinates.) Using the replica trick
and performing a tree-level renormalization-group (RG) calculation
(see, e.g., Ref.\ \cite{cardy-book}), the relevance of $\lambda_{{\cal O}}(\mathbf{x})$
can be gauged by computing the scale dimension $[w_{{\cal O}}]$ of
the second moment $w_{{\cal O}}$ of $\lambda_{{\cal O}}(\mathbf{x})$
at the clean fixed point. If $[w_{{\cal O}}]=d+2z_{0}-2[{\cal O}]>0$,
where $[{\cal O}]$ is the scale dimension of ${\cal O}$ and $z_{0}$
is the dynamical exponent of the clean theory $S_{0}$, then disorder
is perturbatively relevant. This can be viewed as a generalization
of the Harris criterion.

\begin{figure}
\centering{}\includegraphics[clip,width=0.65\columnwidth]{fig1a}\hfill{}\includegraphics[clip,width=0.29\columnwidth]{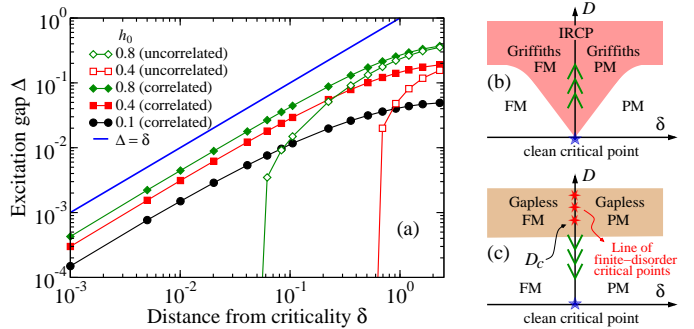}
\caption{\label{fig:gap}(Color online) 
(a) Disorder averaged spectral gap $\Delta$ 
{\it vs}.
distance from criticality $\delta$ for different disorder parameters
$h_{0}$ with and without disorder correlations
(sizes up to $2^{15}$ sites, averaged over $1\ 000$ disorder realizations). 
(b) RG flow diagram
for uncorrelated disorder, the disorder strength renormalizes to infinity.
(c) Flow diagram for correlated disorder ($h_{i}=e^{\delta}J_{i}$),
the clean fixed point is stable. The line of fixed points for strong
disorder can be only accessed for $h_{0}=0$ (see text).}

\end{figure}

Applying these ideas to a $\phi^{4}$ order-parameter theory, ${\cal L}_{0}=r|\phi|^{2}+u|\phi|^{4}+|\partial_{x}\phi|^{2}+|\partial_{\tau}\phi|^{2}$,
we find that quenched disorder can be relevant at tree level only
if it couples to the mass term 
$|\phi|^{2}$.~\footnote{If higher-order terms are disordered, a random mass may be 
created at loop level, but for weak disorder it will be strongly suppressed compared 
to the generic case.}
(We exclude random fields that locally break the symmetry.) This follows
from the fact that $[w_{|\phi|^{2}}]=-d+2/\nu$, which recovers the
original Harris criterion, while disorder coupling to higher order
and gradient terms leads to negative scale dimensions and is thus
irrelevant. Consequently, if the macroscopic random variables do not
produce a random mass term, disorder effects are strongly suppressed.\\
\textit{Transverse field Ising chain---} A tantalizing example
in which the above scenario actually completely removes the random
mass and thus stabilizes the clean critical point even though the
inequality $d\nu<2$ is fulfilled is the ferromagnetic quantum phase
transition of the 1D transverse-field Ising model. Its Hamiltonian
reads 
\begin{equation}
H_{{\rm Ising}}=-\sum_{i}J_{i}\sigma_{i}^{z}\sigma_{i+1}^{z}-\sum_{i}h_{i}\sigma_{i}^{x}~,\label{eq:RTFI}\end{equation}
 where $J_{i}$'s are the interactions, $h_{i}$'s are the transverse
fields, and $\sigma_{i}^{x}$ and $\sigma_{i}^{z}$ are Pauli matrices.
The correlation-length exponent of the clean ($J_{i}\equiv J,h_{i}\equiv h$)
critical point is $\nu=1$, and the Harris criterion predicts that
weak disorder in the $h_{i}$ and $J_{i}$ is relevant. In agreement
with this, {\it uncorrelated disorder} has dramatic effects \cite{mccoy-wu-68,mccoy-prl69,fisher95}.
The clean critical point is unstable against weak disorder and the
RG flows towards an exotic \textit{infinite-randomness critical point}
(IRCP) whose dynamics is so slow that the dynamical exponent $z$
is formally infinite. Surrounding the transition, there are \textit{gapless}
quantum Griffiths phases in which the dynamical exponent $z$ can
be arbitrarily large and the correlation length $\xi$ is finite.
Moreover, the average entanglement entropy ${\cal S}$~\cite{calabrese-cardy-jstatmech04,review-amico-07}
of a subsystem with length $\ell$ embedded in the bulk diverges at
the IRCP as $\overline{{\cal S}}(\ell)\sim(c_{{\rm eff}}/3)\ln\ell,$
where $c_{{\rm eff}}=(1/2)\ln2$ is often called the effective central
charge \cite{refael-moore-prl04,laflorencie-entanglement}. The properties
of this model are summarized in Fig.~\hyperref[fig:gap]{\ref{fig:gap}(b)}.

The order-parameter field theory of this transition can be obtained
from the Hamiltonian (\ref{eq:RTFI}) via a Trotter-Suzuki decomposition
of the partition function and a Hubbard-Stratanovich transformation
of the resulting path integral. After taking the continuum limit,
the resulting $(1+1)$-dimensional action takes the form of a disordered
$\phi^{4}$ theory (see above). The (local) coefficient of the mass
term $|\phi|^{2}$ must be a function of the ratios $h_{i}/J_{i}$
because these are the only two energy scales in the problem. In fact,
it has been shown \cite{pfeuty-pla79} that the critical point of
(\ref{eq:RTFI}) occurs exactly when $\prod J_{i}=\prod h_{i}$, suggesting
that $\delta_{i}=\ln(h_{i}/J_{i})$ is an appropriate measure of the
local distance from criticality~\footnote{At first glance, this definition 
seems to break the symmetry between left and right neighbors in (\ref{eq:RTFI}). 
However, the concept of a distance from criticality is well defined only for 
regions large compared to the lattice constant. 
After averaging $\delta_i$ over such a region, the discrepancy
vanishes.}. Consequently, it is
clear that a simple \textit{local} correlation between $h_{i}$ and
$J_{i}$, say, $h_{i}=e^{\delta}J_{i}$, is sufficient to make the
mass term uniform rather than random.
The disorder of the higher order and gradient terms of the
renormalized action cannot be inferred from this argument, generically
they remain random because they do not depend on the combinations $h_i/J_i$
only~\footnote{It is important to distinguish the bare and renormalized
actions. Pfeuty's criticality condition [16] guarantees that the
{\it renormalized} theory does not have a random mass term. This does not
imply that this term must vanish in any {\it bare} theory.}.
 Although translational symmetry
is broken, the system is locally at the same distance from criticality
everywhere.\\
\textit{Numerical results---} To check the prediction that such
correlated disorder is irrelevant at the clean critical point, we
mapped the Hamiltonian (\ref{eq:RTFI}) onto free fermions using the
Jordan-Wigner transformation \cite{lieb-schultz-mattis,young-rieger-prb96}.
We then performed an exact-diagonalization study contrasting the cases
of uncorrelated and correlated disorder. The absence of Griffiths
phases can be verified by analyzing the gap $\Delta$ in the excitation
spectrum as a function of the distance from criticality $\delta=[\delta_{i}]_{{\rm av}}=[\ln(h_{i}/J_{i})]_{{\rm av}}$
{[}see Fig.\ \hyperref[fig:gap]{\ref{fig:gap}(a)}{]}. Here, the
fields are drawn from a uniform (box) distribution in which $h_{0}<h<1$.
Hence, $h_{0}$ parameterizes the disorder strength. For correlated
disorder, we set $h_{i}=e^{\delta}J_{i}$, while for uncorrelated
disorder, the couplings are independent random variables drawn from
a uniform distribution such that, for the sake of comparison, $h_{0}e^{-\delta}<J_{i}<e^{-\delta}$.
Figure\ \hyperref[fig:gap]{\ref{fig:gap}(a)} clearly shows that
the gap vanishes only at criticality ($\delta=0$) for correlated
disorder, signalling the absence of quantum Griffiths phases. In contrast,
the gap closes before criticality is reached when the disorder is
uncorrelated, a hallmark of quantum Griffiths singularities.

The above uniform distributions of fields and couplings represent
weak or moderate disorder (as long as $h_{0}\ne0$). To study the
fate of the critical point for stronger correlated disorder 
(and, therefore, study nonperturbative effects of disorder not
accomplished by the conventional field-theoretic analysis),
 we now consider a family of gapless power-law distributions 
\begin{equation}
{\cal P}(h)=(1/D)\, h^{-1+1/D},\:{\rm with}\;0<h<1,\label{eq:P}
\end{equation}
 with the bonds given by $J_{i}=e^{-\delta}h_{i}$ as before. $D$
parameterizes the disorder strength. In this case, the system is gapless
for any $\delta$; the absence of a gap simply follows from the low-energy
tail of the distribution ${\cal P}$.

\begin{figure}
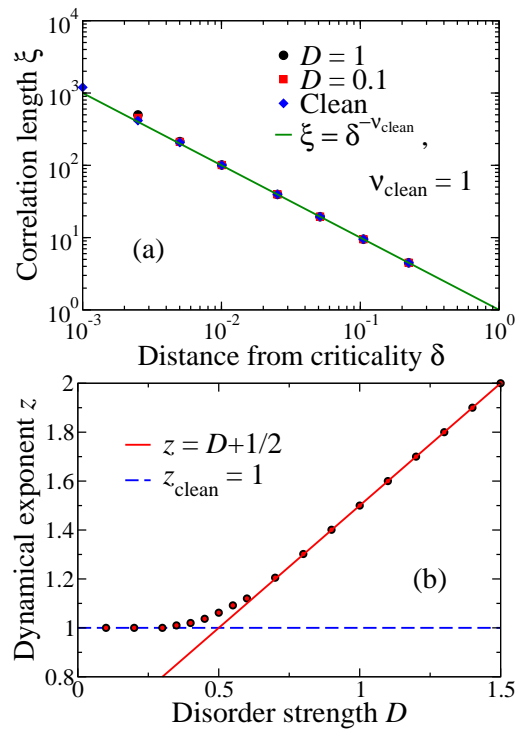

\begin{centering}
\includegraphics[clip,width=0.75\columnwidth]{fig2a}\ 

\includegraphics[clip,width=0.75\columnwidth]{fig2b} 
\par\end{centering}

\caption{\label{fig:nu-and-z}(Color online) (a) Disorder averaged correlation length
$\xi$ as a function of $\delta$
(sizes up to $2\,250$ sites, averages over $20$ disorder realizations).
(b) $z$ \textit{vs.} $D$ at criticality $\delta=0$ (sizes up to
$2^{15}$ sites, typically $1\ 000$ disorder realizations). In both
cases, disorder is correlated, $h_{i}=e^{\delta}J_{i}$, and error
bars are of size of the symbols.}
\end{figure}

We first focus on the correlation length $\xi$ as a function of $\delta$
which is computed by fitting the spin-spin correlation function 
$\overline{\langle\sigma_{i}^{z}\sigma_{i+x}^{z}\rangle}\sim e^{-x/\xi}x^{-\eta}$.
Figure\ \hyperref[fig:nu-and-z]{\ref{fig:nu-and-z}(a)} shows that
$\xi$ takes its clean value regardless of disorder strength $D$.
The correlation length exponent is thus $\nu=\nu_{{\rm clean}}=1$
for all $D$, i.e., it violates the inequality $d\nu>2$ \cite{chayes-etal-prl86}.
This implies that the spin correlations are governed by the {\it clean}
physics analogous to gapped quantum magnets doped with non-magnetic
impurities~\cite{alloul-etal-RMP-09}. We now explore the critical
point $J_{i}=h_{i}$, upon increasing the disorder strength $D$.
The dynamical exponent $z$, computed using the finite-size scaling
of the gap, $\Delta\propto L^{-z}$, is shown in Fig.\ \hyperref[fig:nu-and-z]{\ref{fig:nu-and-z}(b)}.
We see that $z=z_{{\rm clean}}=1$ for $D<D_{c}\approx0.3$, whereas
for stronger disorder, $z$ becomes a monotonically increasing function
of $D$.

The data in Figs.\ \ref{fig:gap} and \ref{fig:nu-and-z} confirm
our prediction that the ferromagnetic quantum phase transition is
governed by the clean fixed point for weak correlated disorder. For
stronger disorder, it is governed by a line of finite-disorder fixed
points as shown in Fig.\ \hyperref[fig:gap]{\ref{fig:gap}(c)}. This
line of fixed points also emerges from an extension of the strong-disorder
RG (SDRG) \cite{fisher95} to correlated disorder \cite{hoyos-unpublished}.
Moreover, for very strong disorder, the result $z\approx D$ can be
understood as arising from the singularity of the bare disorder distribution
via local fluctuations of weakly connected spin clusters. This line
of fixed points belongs to a novel universality class dominated by
disorder effects that are {\it perturbatively} irrelevant. In this
class, some critical exponents (such as $\nu$) are universal and
take their clean values, while others are nonuniversal, such as $z$.\\
 \textit{Entanglement entropy---} We now turn to the ground-state
(GS) entanglement properties at criticality in the case of correlated
disorder, $J_{i}=h_{i}$. We first study the von Neumann entanglement
entropy ${\cal S}(\ell)=-{\rm Tr}(\rho_{{\rm A}}\ln\rho_{{\rm A}})$,
where $\rho_{{\rm A}}$ is the reduced density matrix of a subsystem
${\rm A}$ of size $\ell$. Because we wish to relate ${\cal S}(\ell)$
to local fluctuations of a globally conserved operator, we consider
the spin-1/2 random XX chain 
\begin{equation}
H_{{\rm XX}}=4\sum_{i}t_{i}\left(S_{i}^{x}S_{i+1}^{x}+S_{i}^{y}S_{i+1}^{y}\right)
\label{eq:XX}
\end{equation}
 from now on. It can be mapped onto the Hamiltonian (\ref{eq:RTFI})
by setting $t_{2i-1}=h_{i}$ and $t_{2i}=J_{i}$. The two systems
share the {\it same} entanglement properties at criticality with
$2{\cal S}_{{\rm Ising}}(\ell)={\cal S}_{{\rm XX}}(\ell)$ \cite{igloi-xx-ising-entanglement}.

\begin{figure}
\begin{centering}
\includegraphics[clip,width=1\columnwidth]{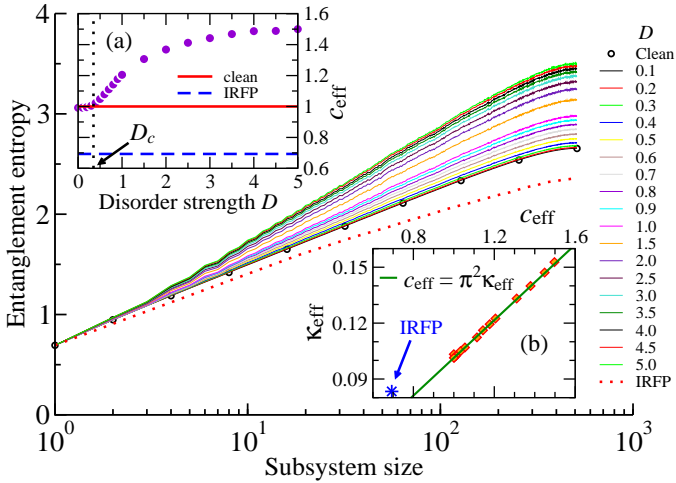} 
\end{centering}

\caption{(Color online) Disorder averaged entanglement entropy $\overline{{\cal S}}$ 
{\it vs.}
subsystem size $\ell$ for the critical GS of (\ref{eq:XX}) with
correlated disorder ($h_{i}=J_{i}$) for various disorder strengths
$D$ (chains of $1\,024$ sites averaged over $5\,000$ disorder realizations).
Inset (a): effective central charge $c_{{\rm eff}}$ {\it vs.} $D$.
Inset (b): magnetization fluctuations amplitude $\kappa_{{\rm eff}}$
{\it vs.} $c_{{\rm eff}}$.\label{fig:entropy}}
\end{figure}

Figure \ref{fig:entropy} shows our numerical results for the average
entanglement entropy $\overline{{\cal S}}(\ell)$, computed via standard
methods \cite{laflorencie-entanglement}. As expected, for weak correlated
disorder, $D<D_{c}$, $\overline{{\cal S}}(\ell)\sim(c_{{\rm eff}}/3)\ln\ell,$
is universal and $c_{{\rm eff}}$ takes the clean value $c_{{\rm eff}}=c=1$.
Surprisingly, for $D>D_{c}$, the entanglement entropy \textit{increases}
monotonically with $D$, i.e., $\overline{{\cal S}}\sim\frac{1}{3}c_{{\rm eff}}\ln\ell$,
with $c_{{\rm eff}}>c$, as shown in inset \hyperref[fig:entropy]{(a)}.
In contrast, for uncorrelated disorder, $c_{{\rm eff}}=c\ln2<c$ \cite{refael-moore-prl04}.
The increase of $\overline{{\cal S}}(\ell)$ with correlated disorder
was first noticed in Ref.\ \cite{binosi-entanglement-prb}, but the
different nature of the GS, its dependence on the disorder strength
and its universality class were not considered. We also point out
that a small deviation from perfect disorder correlations will drive
the system back to the IRCP physics, as revealed by the SDRG and checked
numerically~\cite{hoyos-unpublished}. 

Recently, it has been noted that ${\cal S}(\ell)$ can be related to the local
fluctuations of certain thermodynamic quantities \cite{klich-refael-silva-pra06}.
Here, we use this relation in order to gain further insight on the GS of (\ref{eq:XX}). 
Hence, we compute the fluctuations ${\cal F}_{m}(\ell)$ of the magnetization
$S_{{\rm A}}^{z}=\sum_{i\in{\rm A}}S_{i}^{z}$ of a subsystem A with
length $\ell$: ${\cal F}_{m}(\ell)=\langle\bigl(S_{{\rm A}}^{z}-\langle S_{{\rm A}}^{z}\rangle\bigr)^{2}\rangle$.
In the clean case ($t_{i}=t$), the fluctuations were shown to be
proportional to the von Neumann entropy, ${\cal F}(\ell)=\kappa\ln\ell$,
with $\kappa=\pi^{-2}$, yielding $c=\pi^{2}\kappa$ \cite{song-rachel-lehur-prb10}.
For correlated disorder ($t_{2i-1}=t_{2i}$), ${\cal F}(\ell)=\kappa_{{\rm eff}}\ln\ell$
where $\kappa_{{\rm eff}}$ takes its clean value for $D<D_{c}$ and
increases similarly to $c_{{\rm eff}}$ for $D>D_{c}$, as shown in
inset (b) of Fig.~\ref{fig:entropy}. Interestingly, the ratio $c_{{\rm eff}}/\kappa_{eff}=\pi^{2}$
remains identical to its clean value in the entire $D$-range studied.
Thus even though the line of critical points at $D>D_{c}$ is dominated
by disorder, the relation between fluctuations and entanglement entropy
strongly resembles the delocalized clean system. In contrast, in the
random-singlet state arising for uncorrelated disorder, it is easy
to show that ${\cal F}(\ell)=\kappa_{{\rm eff}}(\ln\ell)$ with $\kappa_{{\rm eff}}=1/12$
\cite{hoyosvieiralaflorenciemiranda}.

Currently, to the best of our knowledge, there is only one general framework in which $c_{{\rm eff}}$ can be understood in terms of random-singlet phases of 1D systems \cite{bonesteel-yang-prl07}. We remark that such framework does not apply to our correlated-disorder case as there is no corresponding random-singlet state. Thus, a fundamental understanding of $c_{{\rm eff}}$ here reported is still lacking and further fundamental understanding of the entanglement entropy in random systems is desirable. Studing the correlated-disorder effects in the many random-singlet states considered in Ref.~\cite{bonesteel-yang-prl07}, and comparing their entanglement entropy with their clean counterparts seems to be a helpful direction of research.\\
 \textit{Experimental realizations---} Consider a quantum $S=1/2$
chain with nearest-neighbor exchange $J_{S-S}$. Doping the chain
with a small concentration of $S_{{\rm imp}}=1/2$ impurities having
$J_{S_{{\rm imp}}-S}\neq J_{S-S}$ would realize a slightly modified
SU(2) (Heisenberg) version of a correlated double-bond Hamiltonian
(\ref{eq:XX}), closely related to the Kondo problem in spin 
chains~\cite{eggert-affleck-prb-92,laflorencie-sorensen-afflec-JSM-08}.
Based on the Matsubara-Matsuda representation of spin-1/2 degrees
of freedom by hard-core bosons~\cite{matsubara-matsuda-PTP-56},
one could also implement the corresponding hard-core bosonic model
(with now correlated random hoppings) using cold-atom systems. Finally,
one-dimensional polymers such as the family of polyaniline \cite{Macdiarmid-87}
are modeled as random-dimer tight-binding chains \cite{wu-phillips-prl91}
which can defy Anderson localization. This random-dimer chain can
be mapped to the Hamiltonian (\ref{eq:XX}). Our results thus provide
an alternative view of the absence of localization in certain 1D electronic
systems.\\
 \textit{Discussion and conclusions---} In summary, we have presented
a general mechanism by which {\it local} correlations between the
random variables render a clean critical point stable against weak
disorder even though it violates the inequality $d\nu>2$. Although
this appears to contradict the Harris criterion, we emphasize that
it merely violates one of its preconditions, namely the spatial
variation of the distance from criticality. Indeed, Harris \cite{harris-jpc74}
argued that a clean critical point is stable if the mean (local) distance
from criticality $[r]_{{\rm av}}$ is larger than the width $\Delta r$
of its distribution. For uncorrelated disorder, this yields 
$\Delta r/[r]_{{\rm av}}\sim\xi^{\frac{1}{\nu}-\frac{d}{2}}<1$
for $\xi\to\infty$, recovering the ($d\nu>2$)-form of the Harris
criterion. However, for our correlated disorder, $\Delta r\equiv0$,
and $[r]_{{\rm av}}\gg\Delta r$ is satisfied regardless of the values
of $d$ and $\nu$.

Our mechanism for suppressing the disorder effects by local correlations
will also operate in $d>1$. The question under what general conditions
the random mass term can be removed completely will be relegated for
future research. Interestingly, Yao {\it et al.} \cite{yao-etal-2010}
recently reported apparent violations of the Harris criterion in several
disordered dimerized spin models. In one system (random dimer model),
this has been attributed to the fact that the quantum critical point
does not depend on the disorder strength, which strongly resembles
our mechanism for the absence of random mass. A similar argument was
given for the Mott-insulator to superfluid transition at the tip of
the Mott lobes \cite{PhysRevB.57.5044}. Another system of Ref.\ \cite{yao-etal-2010}
(random plaquette model) shows a dependence of the critical coupling
on the disorder strength, albeit a very weak one. We emphasize that
in such a case, our mechanism would restrict the deviations from clean
critical behavior to a narrow interval around the critical point that
may well be unobservable.

We acknowledge helpful discussions with David Huse and Ian Affleck.
This work has been supported in part by the NSF under grant Nos. DMR-0339147
and DMR-0906566, Research Corporation, FAPESP, and CNPq. Parts of
this work have been performed at the Kavli Institute for Theoretical
Physics and at the Aspen Center for Physics. NL aknowleges LPT Toulouse
for hospitality.

\bibliographystyle{eplbib}
\bibliography{referencias}

\end{document}